\documentclass[12pt]{svmult}

\usepackage{graphicx}

\usepackage{multicol}

\begin{document}
\title{Rhythms of social interaction: messaging within a massive online network}%
\titlerunning{Rhythms of social interaction}
\author{Scott Golder\and Dennis Wilkinson \and Bernardo Huberman}
\institute{HP Labs, Palo Alto, California}

\maketitle
\begin{abstract}
We have analyzed the fully-anonymized headers of 362 million messages exchanged by
4.2 million users of Facebook, an online social network of
college students, during a 26 month
interval. The data reveal a number of strong daily and weekly
regularities which provide insights into the time use of college students and their social lives, including seasonal variations. We also examined how factors such as school affiliation and informal online ``friend'' lists affect the observed behavior and temporal patterns. Finally, we show that Facebook users appear to be clustered by school with respect to their temporal messaging patterns. 

\end{abstract}


\section{Introduction} 
College students spend a significant amount of time using online social network services for messaging, sharing information, and keeping in touch with one another (e.g. \cite{boyd,Jones}). As these services represent a plentiful source of electronic data, they provide an opportunity to study dynamic patterns of social interactions quickly and exhaustively. In this paper, we study the social network service Facebook, which began in early 2004 in select universities, but grew quickly to encompass a very large number of universities. Studies have shown that, as of 2006, Facebook use is nearly ubiquitous among U. S. college students with over 90\% active participation among undergraduates \cite{Ellison_etal, Stutzman}.

Previous research into Facebook and other social network services, such as Friendster and MySpace, has been performed using surveys (e.g. \cite{Ellison_etal, Stutzman}) and interviews (e.g. \cite{boyd}). While these methods provide a deep understanding of what individuals are doing and their motivations for doing so, they do not capture large-scale patterns or temporal rhythms exhibited by the collective action of immense numbers of users. 

In this paper we present a contrasting view of a social network, one that focuses on the aggregate messaging patterns of over four million members of Facebook. This approach allows us to characterize users' behavior on a large scale. And while personal idiosyncrasies and the massive scale of these social networks might lead one to conclude that one is unlikely to discover any strong global patterns of interaction, our analysis discovered a number of strong regularities across the whole network. Most noteworthy,  messaging within facebook exhibits temporal rhythms that are robust and consistent across campuses and across seasons, resulting from the myriad of individual choices that members make on when and with whom to communicate.  For example, among other insights, the data strongly suggest that college students follow two patterns, a ``weekend'' pattern between midday Friday and midday Sunday, and a ``weekday'' pattern at all other times. Further, our analysis uncovers a grouping effect whereby students in the same university tend to have similar temporal messaging habits, even when the times of day in question do not appear to be a direct effect of the school schedule.

Our large-scale approach also allows us to examine, in a comprehensive manner, the effect of social variables such as school affiliation and online ``friends'' lists on users' propensity to send a message. Nearly all communication was found to occur between ``friends,'' but only a small proportion of ``friends'' exchanged messages. We also found that in messaging there was a slight bias towards members of the same school, while for a particular class of messages known as ``pokes'' the bias was extreme. 

In the remainder of the introduction, we discuss the nature of social networks in the context of the Internet and describe the characteristics of Facebook in the context of previous work. Section \ref{s:data} describes our data set. Results and discussion comprise section \ref{s:results}, followed by a short conclusion. 

\subsection{Social networks and the internet}
A social network consists of all the people -- friends, family and others -- with whom one shares a social relationship. On a macro level, a social network demonstrates how a large group of people are connected to one another. The study of social networks examines both of these levels and, ideally, attempts to relate them to one another. Social network researchers have examined how people make friends and how many friends people have (e.g. \cite{Feld_81,Feld_91}), and how people rely on those in their social networks for social support.

In the past several years, internet access has proliferated, and now internet technologies are useful in supporting relationships and communities, whether proximate or geographically distant. 

The term ``social network website'' describes a class of web services\footnote{For example, Friendster (http://www.friendster.com/), LinkedIn (http://www.linkedin.com/), Orkut (http://www.orkut.com/), MySpace (http://www.myspace.com/), and the one studied here, Facebook (http://www.facebook.com/). This is by no means an exhaustive list.} that invites users to create an online profile of themselves, most commonly with a photograph, a listing of vital statistics (e.g. name, geographic location, sexual preference, occupation) and interests (hobbies, favorite books, movies, television programs, and so on). Most crucially, these services  are focused on allowing users to list other users as ``friends,'' thereby linking their pages to one another and publicly demonstrating their connection. These links between people constitute the ``network'' part of the social network, and enable sharing with friends, including photographs and messages. Often they serve as a way to ``keep tabs'' on people one knows, to keep in touch, or simply to make a list of all the people you can find who you know.

\subsection{Facebook} \label{s:facebook}
Facebook is an online social network website originally designed for college students. Until recently, all users had to have an active student, alumni or work email address from a list of supported academic institutions in order to register\footnote{This created a degree of trust in which, at the very least, users were affiliated with the institutions they purported to be. This is important, given that the visibility of other users' profiles was restricted to those within one's university network. Recently, Facebook opened its doors to anyone who would wish to join; however, school-based networks continue to define social borders and the Facebook culture encourages and enforces users identifying themselves by their real names.}.  The data in our analysis represents a time period and user base comprising almost exclusively college students.

Like other online social network websites, each Facebook user has a personalized profile page which contains personal information and a list of friends. Users may send messages to one another, join topical social groups, and share photographs, weblog posts and brief public messages on a bulletin board called ``the wall''.  Facebook also contains a unique feature called a ``poke'', which is a contentless message.

Facebook users may add other users to the ``friends'' list on their personal profile page. The so-called ``friending'' process involves inviting another user to be one's friend, and the other's acceptance of that invitation. Once friends, a picture and link to the friend's page is added to one's profile page, and vice versa, as friend links are reciprocal and public. By clicking on the links of users' friends, one can navigate through the network of friends. Navigation and browsing through friend lists is a main social activity on the Facebook network. 

As previously mentioned, Facebook use is pervasive among American undergraduates, with over 90\% participation reported from surveys. Since nearly everyone a college student might want to reach (i.e. other college students) can be found within Facebook, it makes Facebook a useful place to communicate with others. It also makes Facebook quite socially relevant, since it becomes the locus for much social interaction; missing out on what takes place in the online world means missing out on a large part of what is happening among one's peers.  It also makes Facebook an interesting research subject. Wellman and Hampton \cite{Wellman_Hampton} suggest that it is once a communication technology is pervasive that it becomes interesting from a research point of view, because that is the point at which it begins to have a real social impact. 

\subsection{Messaging and poking in Facebook}
As mentioned above, users interact socially in Facebook in a variety of ways, including sending private messages and ``pokes,'' which are contentless messages.  

Facebook's messaging capability is similar to that of regular web-based email, except that messages may only be sent to one recipient at a time (distribution lists are not allowed). Messages may be sent to any user, even if the user is not in one's network and even if the sender does not know the recipient's regular email address. Though profiles of users outside of one's own network are not accessible, messaging is one way in which people may have access to others in order to introduce themselves.

Compared to email, Facebook messages are sent infrequently: an average of 0.97 messages per user per week in our dataset. The distribution of messages sent per user has a heavy tail, as discussed in appendix B, which means that a small number of users sent many messages; however, even among those who sent comparatively many messages, the rate of messaging use is smaller than that observed for email \cite{Jones}. Therefore, while the messaging data is too sparse to examine, for example, the messaging habits of individual users, it is nevertheless amenable to studying overall patterns since the aggregate data is quite plentiful.

Pokes may likewise be sent to any one recipient, even if the recipient is not in the sender's network. Pokes appear as a notification, e.g. ``You have been poked by Jane Smith'' on the recipient's login page, inviting a return poke. This ``one bit'' of information, Kaye {\it et al} \cite{Jofish} suggest, is valuable for its open-endedness and ambiguity. Pairs of users are free to ascribe or develop meaning for the poke that is unique to their relationship, or even to poking in a certain context. ``Poke wars'' are somewhat common anecdotally, as well as in our data - a pair of users repeatedly poke one another back and forth over a period of hours or days\footnote{While most poke wars comprise a valid social interaction, automated scripts do exist that allow users to send a very large number of pokes in rapid succession. We controlled for these bots as described in Appendix A.}. Kaye {\it et al} \cite{Jofish} observed in the use of a ``virtual intimate object'' (a kind of desktop-based poking tool for romantic couples) an obligation to reciprocate such that ``clickwars'' developed, and notes the same kind of behavior in literature on text messaging. We observed similar behavior in Facebook.

\subsection{Messaging and poking as proxies for online social activity}\label{s:argue}
While measuring the number of messages and pokes exchanged by college students can be interesting in and of itself, we further suggest that messaging and poking serve as proxies for gauging and understanding online social activity on a large scale. Email exchange has long been used as a proxy for measuring the strength of relationships (e.g. \cite{Tyler_Wilkinson}). Sending a message or poke is a discrete event that represents an active, socially meaningful gesture by the sender. Further, since messaging is private, it is less subject to the pressures of self-presentation than affect other online social networking capabilities such as friend selection and profile items. 

Granovetter \cite{Granovetter} notes that one of the measures of the strength of a relationship is the time and effort invested in maintaining it.  Interacting through messages and pokes certainly represents an investment in maintenance, in contrast to friend links, which are eternal and do not require any effort or upkeep.  Friend links are certainly of interest, but because they are established by fiat, rather than through regular interaction, it is difficult to examine such a network and separate what is effectively users' self-reported behavior from their true feelings and motives, as evinced by their actions\footnote{For example, anecdotally, users have a very low threshold for accepting friend requests, often accepting requests from acquaintances or even strangers, perhaps in order to avoid hurt feelings.}.  One goal of this paper is in fact to determine the extent to which the ``friends'' network is similar to the network created through the regular course of user interaction through messaging, and this topic is addressed in section \ref{s:whofriend}.

\subsection{Time spent communicating online}
Our investigation into temporal patterns in Facebook takes into consideration previous work in identifying why and when college students use electronic communication tools, as well as computers more generally.

Grinter and Palen \cite{Grinter_Palen} studied the use of instant messaging (IM) among teens and college-age people. Like Facebook, instant messaging is a popular method of communication among this age group. Grinter and Palen found that time constraints affected the use of IM. For example, teens were subject to the temporal constraints of the home, e.g. dinnertime, but largely had the same schedule as other teens. This makes engaging in synchronous communication much easier. While college students had less predictable schedules, their schedules were also less likely to overlap with one another. Nevertheless, instant messaging while multitasking was a prevalent feature, and when doing homework, etc., it would be possible to IM simultaneously.

It is clear from the Grinter and Palen study, as well as from others (e.g. \cite{Jones}) that people in this age group spend a great deal of time using computers, with online communication being either a primary or secondary use. In a way, Facebook time and IM time represent computer time, because they are activities that are engaged in parallel with other computing activities.

How college students spend their time is of great concern. Their physical, academic and social well-being is affected by the choices they make, and making healthy choices can be challenging when it is one's first time having the freedom and responsibility to make the majority of those choices for oneself. Students' academic performance is predicted by how much time they spend studying \cite{Stinebrickner} and where students spend their time and with whom they spend it predicts whether they will binge drink \cite{Clapp_etal}, for example.

Our contribution to the question of how college students spend their time will consist of an understanding of when they are using computers as social tools, and with whom they're communicating, and when. We cannot say with any certainty what else they are doing at that time, or what they are doing with the rest of their time, but we demonstrate that regular ebbs and flows to their computer time characterize their daily lives.

\section{Data} \label{s:data}
Our data set consisted of fully anonymized headers, with no message content, for messages and pokes sent by 4.2 million users of Facebook who were members of one of 496 North American colleges and universities. The set included the 284 million messages and 79.6 million pokes sent by these users between February 2004, when Facebook was created, and the end of March 2006.  The message and poke timestamps were normalized to the local (school) time zone.

The dataset also included anonymized friends lists for each user, current as of March 31, 2006. The friends list did not include timestamps indicating when the two people became friends. Therefore, in our analysis, when we say that a message was sent between friends, we mean that the individuals were friends by the time our dataset was created; the individuals may or may not have been friends by the time a particular message was sent. This limitiation of the dataset made it impossible to determine which communications preceded the establishment of friendship, and which came after.

Before we acquired the data, each user was assigned a randomized 10-digit ID number, while each of the 496 universities was assigned a randomized ID number between 1 and 500. The users were grouped by school. The message and poke data contained only the sender's and recipient's randomized user IDs and a timestamp. The friends data contained the randomized user IDs corresponding to each user's friends list. 

Before further processing, we cleaned the data to remove spam and junk, as described in appendix \ref{s:A}. This process removed 43.0\% of the original message data and 0.402\% of the original poke data. The total number of messages in the cleaned data set was 162 million, while the number of pokes in the cleaned set was 79.2 million, for a total of 241 million non-junk communications. The total number of friend links was 378 million.

Our dataset was drawn from college networks, and surveys \cite{Stutzman, Acquisti_Gross} have shown  that undergraduates make up the overwhelming share of Facebook users in these networks. Further, we presume that there is a strong cohort effect in place, and given that Facebook has been popular since 2004 this also suggests that the non-undergraduate population is comparatively small. Our results are thus highly relevant to undergraduates alone.

\section{Results} \label{s:results}
We begin by discussing overall messaging trends in relation to school affiliation and ``friends'' lists. Next, we explore the temporal rhythms of messaging, identifying strong weekly and daily regularities both in the aggregate data and particular subsets. We examine seasonal variation in temporal messaging patterns, and finally identify correlations between temporal patterns and school affiliation.

\subsection{Who is a friend?} \label{s:whofriend}
Messages and pokes were largely exchanged between people who have listed one another as friends. In our data, 90.6\% of messages and 87.5\% percent of pokes were exchanged between friends. Though a large proportion of messages were sent to friends, it is emphatically not the case that most friends were message recipients. On the contrary, of 378 million friend links, only 57.0 million (15.1\%) of those friend pairs exchanged messages.

\begin{figure}[h!]
    \centering
    \scalebox{0.60}[0.60]{\includegraphics[viewport= 100 475 500 750]{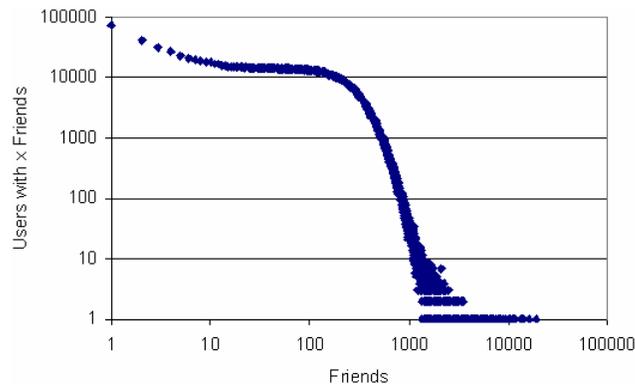}}
    \caption{Distribution of number of friends per user in Facebook}
    \label{f:frdist}
\end{figure}

Of the 4.2 million users in our dataset, we found a median of 144 friends and mean of 179.53 friends per user. This finding is in agreement with survey results from Ellison et al. \cite{Ellison_etal}. The difference between mean and median was caused by a small number of members with a very large number of friends (e.g. 11 users with more than 10,000 friends) each.  It may be of interest to some to compare this result to the speculative ``theoretical'' limit of 150 friends given in \cite{Dunbar}.  Figure \ref{f:frdist} shows the degree distribution of friend links. 
On the $x$ axis is the number of friends, and on the $y$ axis the number of people with that number of friends. We see that, to the left, thousands upon thousands of people have anywhere from 1 to a few hundred friends, but at about 250 friends, the number starts to drop sharply. 

This finding underscores the problematic nature of ``friend''-ship. As we discussed previously, people add friend links for a variety of reasons, not always for reasons that imply the pair are friends in the conventional sense, that they interact socially and share a mutually important connection of trust, affection, shared interests, and so on. Of course, communication in Facebook cannot possibly be shown to represent the total communication between a pair of individuals, since out-of-network emails, instant messages, telephone calls, and face-to-face interaction are not represented. But while absence of messaging within a friend pair is not evidence of lack of a bond, the existence of messaging does constitute positive evidence of the existence of a bond. Because so many people are listed as friends as compared to those who engage in active messaging, for research purposes, being a friend in Facebook can be considered a necessary but not a sufficient condition for being a friend in the conventional sense.

\subsection{School ties}
Facebook attempts to be a community that is, at least, partially geographically-bounded. Users are each associated with a particular university network, and profiles are visible only to those within that network, unless the individuals are friends. This feature has the desirable property that one can remain semi-private, with respect to the outside world, yet still be accessible. Ellison et al. \cite{Ellison_etal} showed the importance of Facebook in fostering several kinds of social capital on a single campus. While the importance of Facebook within a school is demonstrably high, Facebook also serves as a keeping-in-touch communication tool, much like the telephone, instant message and email. While about half of all messages are sent to friends in one's own school, 41.6\% of messages are sent to friends in different schools (table \ref{t:mess}).  
\begin{table}
\centering
\begin{tabular}{|c|c|c|} \hline
\% messages & Same school & different school \\ \hline
Friends & 49.0 & 41.6 \\ \hline
Nonfriends & 5.9 & 3.5 \\  \hline
\end{tabular}
\caption{Proportion of messages by recipient type} \label{t:mess}
\end{table}
Note that the differences between percentages in tables \ref{t:mess} through 4 are statistically significant by a large margin because of the huge number of data involved.

These cross-school links are important, as they represent pairs of people whose bonds are not a result of meeting at the university. Feld \cite{Feld_81} observes that a relationship has a focus, or a shared circumstance around which interaction takes place. A focus might be a geographic community, an academic institution, social or interest group, family, or other reason that people group together. In Facebook, the primary network is often the focus around which the relationship was formed, but this is not the case for friend pairs from different schools. Given that different-school friend pairs represent such a large proportion of the messages sent, we suggest that keeping up with distant friends is one of the main ways Facebook is used for social interaction.

Other kinds of foci may be represented in Facebook in data that was not part of our study. For example, college students often link themselves to their high schools, the geographic regions in which they live, summer programs they attended, and so on. Such listings might shed some light on the foci through which pairs of friends met, but the college network remains the one most likely to be central in the lives of college students at the time they are active in Facebook.

But what about messaging pairs of people who are not friends? Given the rather low cultural barrier to adding friends, we must conclude that these are people who are not interested in knowing one another further. Also, the proportion of messages sent to non-friends in the same school is greater than the proportion sent to friends. Either people are more likely to eventually list people from a different school as friends, which makes little sense, or people are more likely to send messages to non-friends if they are in the same school.

Like messages, pokes are predominantly sent to people in the same school as compared to people in different schools, and to friends as compared to non-friends.  The proportion of pokes sent to friends (87.6\%)  is very close to the proportion of messages sent to friends (90.6\%). However, the bias toward people in the same school is quite extreme for pokes.  While 54.9\% of messages were sent to recipients in the same school, demonstrating a slight bias, an overwhelming 98.3\% of pokes were sent to people in the same school (table \ref{t:pokes}). 
\begin{table}
\centering
\begin{tabular}{|c|c|c|}\hline
\% pokes & Same school & different school \\ \hline
Friends & 86.6 & 0.97 \\ \hline
Nonfriends & 11.72 & 0.72 \\ \hline
\end{tabular}
\caption{Proportion of pokes by recipient type} \label{t:pokes}
\end{table}
That so few pokes are sent to people at different schools is surprising finding, one for which we cannot provide an explanation.  Though research like Kaye \cite{Jofish} suggests that remote presence awareness is valuable to users who are separated geographically, the lack of pokes to those people implies otherwise.

\subsection{Reciprocity}
To further complicate the use of messaging as a
proxy for social interaction, we recognize that
people often receive messages that are unwanted. In
conventional email, spam plagues most email users,
and even friends who send too many jokes and chain
letters can be an annoyance. It would not be fair to
characterize a relationship as existing when the
message recipient is an unwilling participant. We
examined which sender-recipient pairs had
reciprocated relationships; that is, whether each
partner in the exchange was both a sender and a
recipient of messages to the other. 
From the results below, we conclude that both being in the
same school and having being linked as friends are
indicative of the existence of a social relationship
that increases the incidence of message exchange.

As we noted above, being in the same school makes it
easier to send messages to non-friends; it turns out
that it also affects whether the individuals message
one another reciprocally.  When a user sends
messages to someone in the same school, that person
(eventually) messages them 58.8\% of the time.  But
when the people are in different schools, this
happens only 40.8\% of the time.  Given the very
large number of datapoints, this difference is
statistically significant. Likewise, having an
established friend link significantly increases the
likelihood of reciprocal messaging.  When two
individuals are friends, messaging is reciprocal
51.7\% of the time, compared to 42.6\% of the time
when the two are not friends.

Next, we look not at individual users' messaging but
rather at pairs of users who communicate with one
another.  The aforementioned effects are similarly observed when user pairs are considered.  Controlling for both friend status
and school affiliation, we obtain the data shown in
table \ref{t:recippairs}.
\begin{table}
\centering
\begin{tabular}{|c|c|c|}\hline
\% pairs & Same school & different school \\ \hline
Friends & 43.2 & 26.6 \\ \hline
Nonfriends & 33.2 & 18.1 \\ \hline
\end{tabular}
\caption{Proportion of message pairs that are
reciprocal, controlling for friendship and school} \label{t:recippairs}
\end{table}

It might come as no surprise that relationships in
which messaging is reciprocated account for a
disproportionate amount of messages, since the
positive feedback loop created by mutual response
would be expected to prompt future interactions.  In
fact, this is the case in every category (friends,
same school, etc.), as shown in the table \ref{t:recipprop}.
\begin{table}
\centering
\begin{tabular}{|c|c|c|}\hline
\% messages & Same school & different school \\ \hline
Friends & 72.3 & 48.1 \\ \hline
Nonfriends & 58.0 & 34.3 \\ \hline
\end{tabular}
\caption{Proportion of are messages that are between
reciprocal messaging partners, controlling for
friendship and school} \label{t:recipprop}
\end{table}

But in reciprocal relationships, two people are
sending messages rather than one, so we must return
to looking at individual senders, rather than pairs.
To a partner who (eventually) sends a message back,
the average individual sends an average of 2.29
messages, whereas with a partner who never responds,
the average is 1.57 messages. Perhaps users
give up after sending one to two messages and never
receiving a response.

\subsection{Temporal rhythms}
The temporal patterns of messaging and poking in Facebook display strong weekly and daily regularities. Since (as discussed above in section \ref{s:argue}) messaging\footnote{We experimented with other such measures, such as counting the number of unique users sending messages, rather than the messages themselves, but in all cases the same trends obtained, but of course with different raw counts. Likewise, the trends for messages and pokes were nearly identical, but with different counts.} is a proxy for all online social activity, these results are quite useful for illuminating the social and computer time use of college students.

As a preliminary note, we found that temporal patterns for poking are in almost all cases indistinguishable from messaging. Thus, unless otherwise noted, the results below are for messages and pokes combined.

\begin{figure}[h!]
    \centering
    \scalebox{0.4}[0.4]{\includegraphics{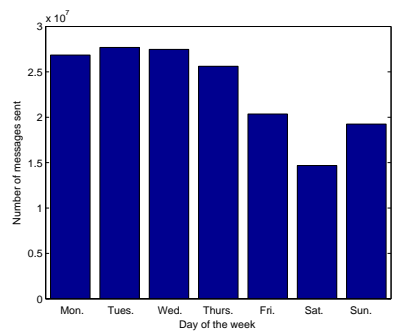}}
    \caption{Message plus pokes sent by day of the week}
    \label{f:bardays}
\end{figure}
We begin by exploring temporal patterns in Facebook by observing how messaging and poking varies on a day-to-day basis. Figure \ref{f:bardays} shows that use is at its highest on the first few days of the week. By Thursday, use begins to decrease, and use is lowest on the weekend. This in itself is interesting, but the picture becomes clearer once it is broken down into hours as well as days.

\begin{figure}[h!]
    \centering
    \scalebox{0.75}[0.70]{\includegraphics[viewport= 100 250 500 550]{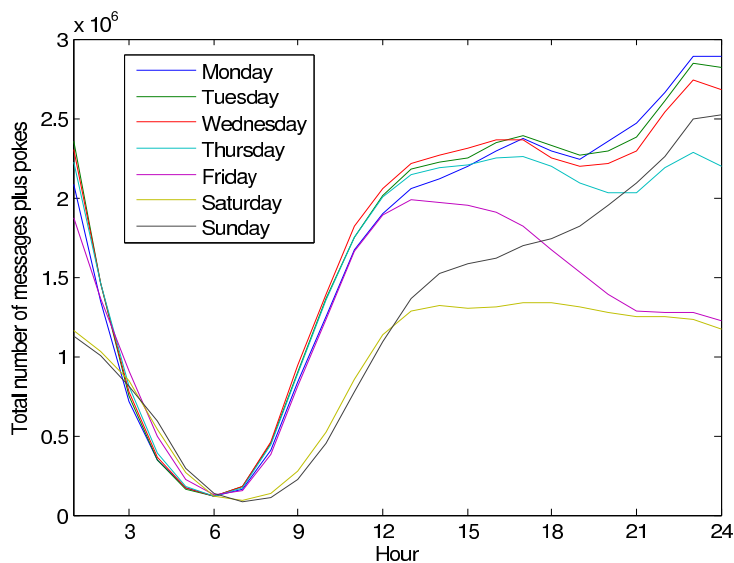} }
    \caption{Message plus pokes sent by hour in Facebook (color)}
    \label{f:week}
\end{figure}
Figure \ref{f:week} (color) shows the total number of messages plus pokes sent within each hour of each day of the week.  Each of the seven lines represents a day of the week.  This figure aggregates data from all schools over the entire 26 month period. In fact, the same pattern is observed (both ``by eye'' as well as using statistical techniques, as described in appendix \ref{s:C}) for single schools considered alone, and also over shorter time periods such as weeks or months. 
 
These messaging patterns suggest that the college student weekend, rather than consisting of solely Saturday and Sunday, may be considered to run from mid-Friday to mid-Sunday. We may say that Monday, Tuesday, Wednesday and Thursday are complete weekdays; they each exhibit patterns nearly identical to one another, with little messaging activity between roughly 3 and 8 am, and increasing messaging activity throughout the day until the evening, when there is a dip in activity, until steadily rising until roughly midnight.

By contrast, Saturday is a pure weekend day. Messaging activity is lesser overall, and is flat, rather than increasing, in the night. Friday and Sunday can be considered hybrid weekday-weekend days. Friday tracks the other weekdays until about 1:00 pm, at which point activity sharply decreases, and from 9 pm onward, it tracks Saturday. Conversely, Sunday tracks Saturday, also until about 1 pm, at which point activity steadily rises, and by 9 pm it resembles weekdays.

In section \ref{s:argue} we argued that messaging patterns are a reliable proxy for activity. We demonstrate this here by way of comparison to messaging in a corporate network.
\begin{figure}[h!]
    \centering
    \scalebox{0.55}[0.45]{\includegraphics[viewport= 100 250 500 550]{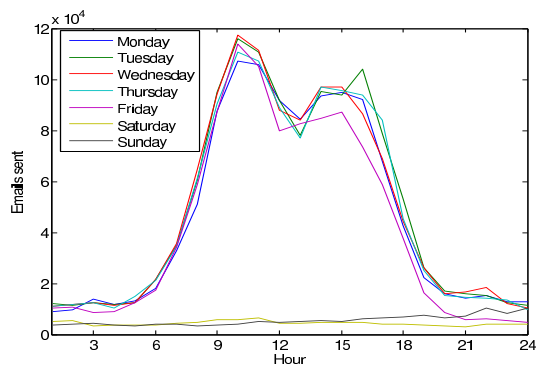} }
    \caption{Message plus pokes sent by hour in a corporate network (color)}
    \label{f:corp}
\end{figure}
Figure \ref{f:corp} (color) shows messaging activity by hour in a corporate email network, using the same dataset as used in \cite{Tyler_Wilkinson}. The familiar pattern of the work day is observed in the data: from Monday to Friday, email activity begins rapidly around 7 to 8 am and decreases rapidly around 6 pm, with an early afternoon decrease around lunchtime,  while Saturday and Sunday show very little activity.

Messaging patterns thus encapsulate the differences between college students' lives and those of employees of a corporation.  While people in the working world have a five day schedule characterized by what are conventionally known as business hours, college students have a schedule in which they integrate computer use into most of their waking hours.

Facebook use is at its lowest during the college student
weekend, presumably when students are away from
their computers, especially Friday and Saturday
nights, culturally seen as time for
socializing.  The conclusion is that Facebook use,
and therefore computer use, does not represent
leisure time, but rather social interaction engaged
in as an activity paralleling the schoolwork and
other computer-related activities during the week.
This is significant, in that it represents
large-scale quantitative evidence that supports
claims about student messaging and internet behavior
as well as student behavior regarding studying and
socializing. If online communication and internet
use are multitasking activities alongside schoolwork
\cite{Grinter_Palen,Jones}, then the points at
which Facebook use is high would represent the time
the most schoolwork is done.  Indeed, it would
explain why internet use increases with time spent
studying \cite{Korgen_etal}. The complement of this
is that when students are socializing rather than
studying, Facebook use would be lowest.  This
appears to be true; alcohol consumption is a social
activity among college students, which tends to take
place primarily on the weekends \cite{Clapp_etal},
which is when Facebook use is lowest. Furthermore,
this pattern supports the ``displacement'' model of
internet use proposed by Nie and Hillygus \cite{Nie_Hillygus},
which hypothesizes that internet use during the
weekend leads to a decrease in sociability; Facebook
users appear to prefer offline socializing to online
socializing during the weekend.

\subsection{School and friend ties over time}
\begin{figure}[ht!]
	\centering
	\scalebox{0.65}[0.40]{\includegraphics{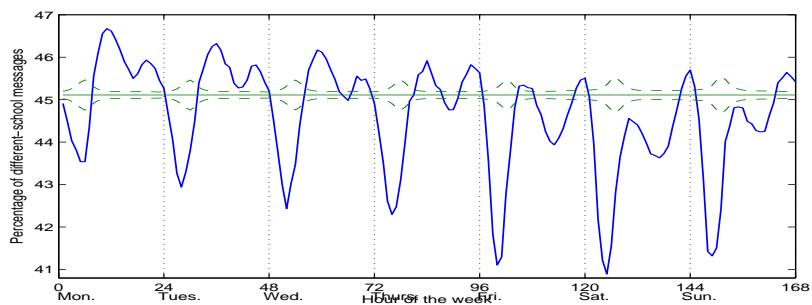} }
	\caption{Messages to recipients in a different school by hour of the week; the three green lines show the	overall average and the average plus and minus two standard deviations. The standard deviation varies because of the varying number of sent messages by hour in the data.}
	\label{f:frdiffsch}
\end{figure}

The proportions of messages and pokes sent to friends and same-school recipients also exhibit consistent patterns over the week.  In general, the temporal patterns in the data remain stable over different subsets of messages or pokes. For example, the percentage of messages plus pokes sent between members of different schools remains close to the overall average of 45\% throughout the week, as demonstrated in figure \ref{f:frdiffsch}.

\begin{figure}[h!]
	\centering
	\scalebox{0.65}[0.40]{\includegraphics{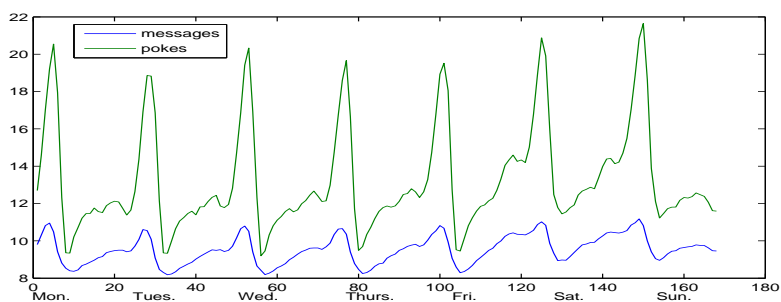} }
	\caption{Percentage of pokes (upper curve, red) and messages (lower curve, blue) to nonfriends by hour of the week. The black lines show, respectively for each plot, the average and the average plus and minus two standard deviations.}
	\label{f:nonfr}
\end{figure}

While small, however, the deviations from the average shown in figure \ref{f:frdiffsch} and the others in this section display not just statistical significance, but more interestingly, a strong weekly rhythm. For example, the percentage of messages sent to recipients at different schools (figure \ref{f:frdiffsch}) is at its highest during weekday daytime, and sharply plummets in the late night hours. Similarly,  the percentage of messages and pokes sent to nonfriends at the same school (figure \ref{f:nonfr}) exhibits a sharp late-night spike. Both these trends hold for all days, even weekends.

\subsection{Seasonal variation}
One might expect that students' weekly and daily messaging patterns change significantly during the summer as many of their daily schedules no doubt change. In fact, this is not the case. The overall weekly pattern during, for example, July is remarkably (and statistically in the sense of appendix \ref{s:C1}) similar to the overall average shown in figure \ref{f:week}. Some minor differences may be observed, such as an unaccoutable increase in activity on July Mondays, but such anomalies are observed in every month (e.g., Friday and Saturday night decreases in October). No trend that we observed was consistent with a significant seasonal shift in daily or weekly temporal pattern, even when pokes or messages are considered separately.

\begin{figure}[ht!]
	\centering
	\scalebox{0.55}[0.45]{\includegraphics{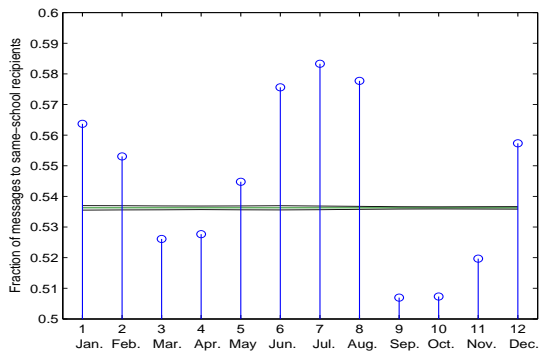} }
	\caption{Fraction of messages sent to recipients in the same school in 2005, by month (blue circles); the green and black lines show the overall average and the average plus or minus two standard deviations, respectively}
	\label{f:monthly}
\end{figure}

While the temporal rhythms of messaging do not change from month to month, a statistically significant variation exists in the fraction of messages sent to same-school correspondents. This is shown in figure \ref{f:monthly}. Note that only the first 28 days of each month were considered so that each point corresponds to four full weeks.
The proportion of messages sent to same-school recipients increases dramatically, in comparison to the standard deviation, during June, July, August, December and January.  These months correspond to summer vacation and the Christmas/winter break between semesters.  These are the times when most students are not on campus, which suggests a simple explanation: since other communication channels, i.e. face-to-face, are eroded due to distance, messaging becomes more heavily relied upon to maintain contact.

\subsection{Variation by school: clustering effect}
In this section we show that the data suggest that students are grouped together by school according to their temporal messaging pattern. As an example, we observed that schools with higher than normal activity during weekend daytime hours are extremely likely to have higher than normal activity during weekday late-night hours. This correlation and many others are elucidated below. 

\begin{figure}[h!]
    \centering
    \scalebox{0.7}[0.4]{\includegraphics{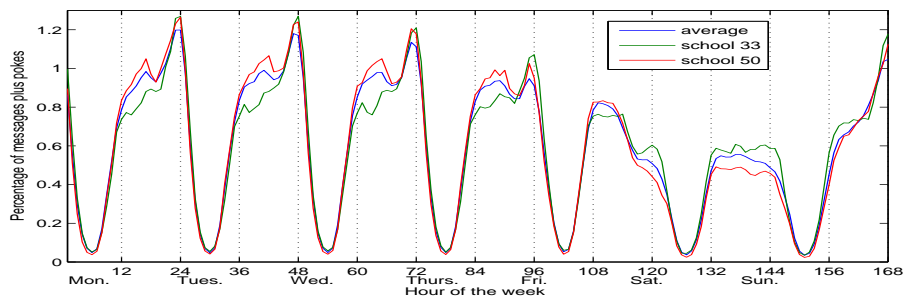} }
    \caption{Deviations from the average weekly pattern for two schools (color)}
    \label{f:2dev}
\end{figure}

We first consider for each school how much its messaging activity (that is, number of messages sent) deviates from the overall average weekly pattern of \ref{f:week}. As an example of deviations from the overall pattern, consider figure \ref{f:2dev} displaying the weekly pattern of schools 33 and 50 compared to the overall average.
The figure shows that users from school 33 sent fewer messages than average on weekday mornings and more than average on weekday evenings and weekends, while users from school 50 sent more messages than average on weekday morning and evenings but less than average on weekends.  

While individual schools' deviations might be of some interest, it is the consistent and pervasive nature of these deviations which is particularly intriguing.  For example, our data indicate that schools which have above average activity during the Monday 0--1 hour\footnote{That is, 12 midnight to 1 AM. The notation 0--24 for the hours of each day is used throughout what follows.} are very consistently above average for the Tuesday 0--1 hour, while schools which are below average for Monday 0--1 are very consistently below average for Tuesday 0--1. We thus say that the hours 0--1 Monday and 0--1 AM Tuesday are strongly correlated across different schools.

Were students divided randomly into groups, such a correlation would not be expected to occur. Instead, deviations from the average would be small and would not demonstrate correlation. Several explanations are possible for the correlations we observe, including school schedule. However, as we demonstrate below, school schedule alone is not enough to explain the observed correlations.

The correlation between two hours of the week, say hours $x$ and $y$, may be described mathematically by way of the correlation coefficient $r_{xy}$. The meaning of $r_{xy}$ is as follows: a large positive or negative value\footnote{in practice the coefficients are normalized so that the highest value is +1 and the smallest -1} of $r_{xy}$ indicates that schools' deviations from the average for hours $x$ and $y$ follow a consistent pattern. Conversely, a zero or small value indicates that schools' deviations for hours $x$ and $y$ do not follow a consistent pattern. A large positive value means that the counts for these hours tend to be both high or both low, while a large negative value means that the deviations tend to occur in opposite senses. The mathematical details of obtaining this coefficient are described in appendix \ref{s:C2}. 

\begin{figure}[h!]
    \centering
    \scalebox{0.70}[0.70]{\includegraphics{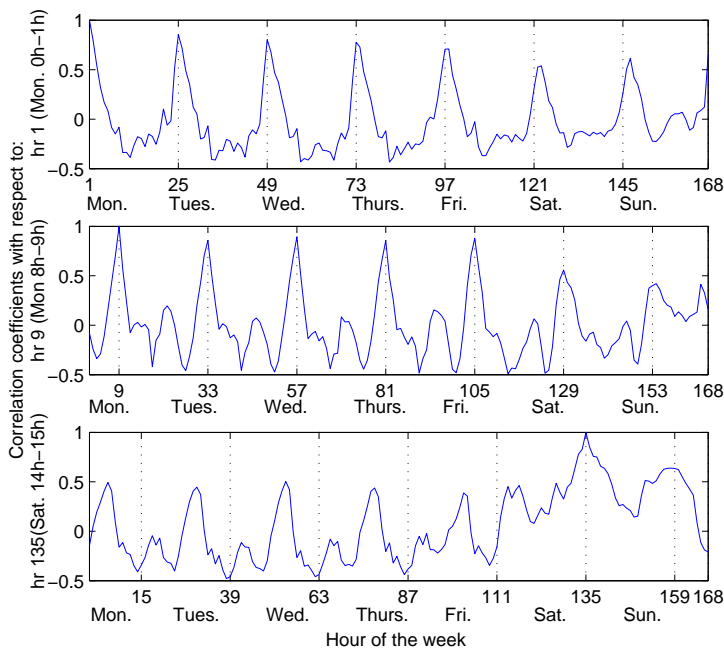} }
    \caption{Correlations with respect to hours 1, 9 and 135 of the week}
    \label{f:r3}
\end{figure}

\begin{figure}[ht!]
    \centering
    \scalebox{0.60}[0.60]{\includegraphics{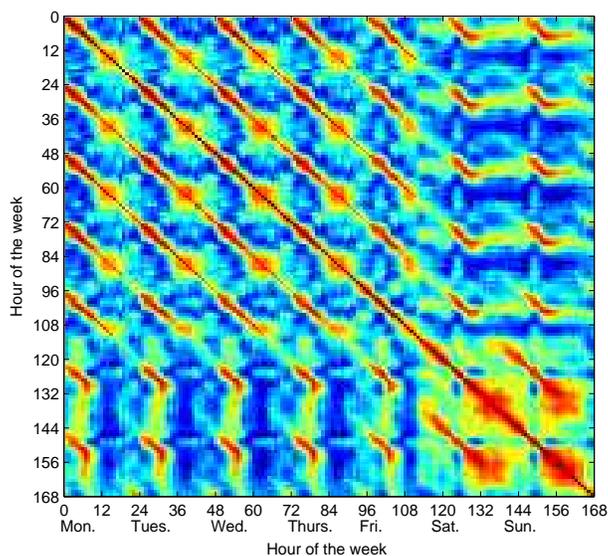} }
    \caption{Matrix of correlations between hours of the week; red = positive, dark blue = negative, light blue = no correlation}
    \label{f:r}
\end{figure}

Correlation coefficients are presented in figures \ref{f:r3} and \ref{f:r}. Figure \ref{f:r3} shows correlations of all hours of the week to three reference hours: hour 1, a late night hour (Monday 0--1); hour 9, a morning hour (Monday 8--9); and hour 135, a weekend hour (Saturday 14--15).  Figure \ref{f:r} (color) shows the (symmetric) matrix of correlations for every pair of hours of the week. The very strong correlation between most periods separated by exactly 24 hours (e.g., Tuesday 8--9 and Wednesday 8--9) is perhaps to be expected.  Other notable features include the strong positive correlation between weekday morning hours and other weekday morning hours, late-night hours and other late-night hours, and weekend daytime hours with other weekend daytime hours. Negative correlations are most notably observed between late-night and weekday daytime hours

Other, more intriguing features of figures \ref{f:r3} and \ref{f:r} include the previously mentioned positive correlation between late-night and weekend daytime hours, and the negative correlations between weekend afternoon and weekday afternoon hours. Hours for which no correlation or only a small correlation is observed may also be of interest, such as weekday 20--21 (a peak of activity) and weekday daytime hours, which are almost totally uncorrelated.
 
For certain hours of the week, a strongly positive or negative correlation over different schools is perhaps to be expected. The level of activity during the weekday morning hours, for example, may determined by school schedule; users at schools which schedule more (or fewer) classes in the morning might be expected to demonstrate increased (or decreased) activity levels across the weekday mornings, as observed. This could also account for the negative correlation between weekday late-night and weekday midday hours.

However, for many other hours, the correlation cannot be accounted for by schools' administrative decisions alone. One example is the correlation between weekend hours and other weekend hours. Apparently, different universities consistently contain either a disproportionately large or disproportionately small number of weekend-active Facebook users. The origin of this trend presents an interesting research question.

\section{Conclusion}
Facebook is, for the time being, a dominant locus for college students' electronic
social activity.  Use of Facebook is weaved into the college student experience, and its use mirrors
college students' daily, weekly and seasonal
schedules.  Its value lies in its use as a way for
college students to support both distant and
geographically proximate relationships.

In this paper we examined patterns and trends among 362 million anonymized messages and ''pokes'' sent by 4.2 million Facebook users.  We found a strong weekly temporal pattern to college
students' Facebook use, a grouping of students with similar temporal patterns 
by school, and a seasonal variation in the
proportion of messages sent within a school.  Our
study further revealed that messages are mostly sent
to friends, but most friends do not receive
messages, demonstrating the problematic status of
the ``friend'' link and the value of messages over
friend links for studying online social network
systems. 

{\bf Temporal Rhythms Overview} That
a college student ``weekend'' is so clearly visible
in the data lends strong support to other research as discussed.
Electronic communication takes place alongside
schoolwork, which is largely computer-based, and
gives an explanation why time college students spend
using the internet increases with their study time.  It also is consonant with the
``displacement'' model of internet use that suggests weekend time spent on
the internet decreases sociability, and shows that
students indeed are not using Facebook -- and
therefore, likely, the internet -- as much on
weekends. Seasonal variation in
same-school/different-school messaging proportions
confirms that messaging is used in support of
geographically distant relationships.  We showed
that messages within one's school are higher
precisely at the times students are not at school,
which suggests that there is a greater likelihood to
send messages to people who are not close by.

{\bf Messages Versus Friends} Most messages
are sent to friends.  However, most friend pairs do
not exchange messages, suggesting it's easier to
have lots of friends than lots of message partners. 
Since messaging requires an investment of time and
energy on the part of the sender, it evinces social
interaction in a way that friend links do not.  We
therefore propose that messaging is a more reliable
measure of Facebook activity.

\subsection*{Acknowledgments}
Thanks to Adam D'Angelo, Wayne Chang and Jeff Hammerbacher from Facebook for their assistance in providing the data analyzed in this paper.

\appendix

\vspace{0.15in}

{\bf \noindent \Large Appendices}

\vspace{-0.25in}

\section{Spam and junk in Facebook} \label{s:A}
To eliminate spam or junk messages from consideration, we cleaned the message and poke data in the course of our analysis. Examples of spam messages include the 19946 messages sent by user 568592864 to one other user in one 42 second period, or the 8634 messages from user 149676784 to various other users over one 31 second period. As mentioned in the text, such sets of messages were relatively common, comprising 43\% of all messages. Some users even sent batches of pokes (which contain no content!), although there were far fewer spam pokes than spam messages. 

The large batches of messages were not primarily messages to distribution lists, because Facebook did not support this feature except during a small period of time and almost all of the messages we identified as spam fell outside of this time period. Even if a small number of messages to distribution lists were eliminated as spam, this does not affect our analysis since sending to large distribution lists is not a good proxy for social interaction. Nor did the batches of pokes represent ``poke wars'' as discussed in the text, because the pokes in the batches all went to different recipients\footnote{In fact, Facebook does not allow users to poke a recipient a second time until the recipient has either responded to or dismissed the first poke.}.

Analysis indicated that the time separation between the messages or pokes in large batches was always lower than 5 seconds. To clean the message data, we simply removed any message sent by a user within 5 seconds of his previous message. It is possible that this removed very small numbers of legitimate messages which should not affect the analysis. To clean the pokes, we removed only batches which contained more than 20 pokes separated by less than 5 seconds, because it is conceivable that users could exchange a small number of pokes in quick succession.

\section{Distribution of number messages sent per user}\label{s:B}
\begin{figure}[h!]
	\centering
	\scalebox{0.43}[0.43]{\includegraphics[viewport= 100 250 500 550]{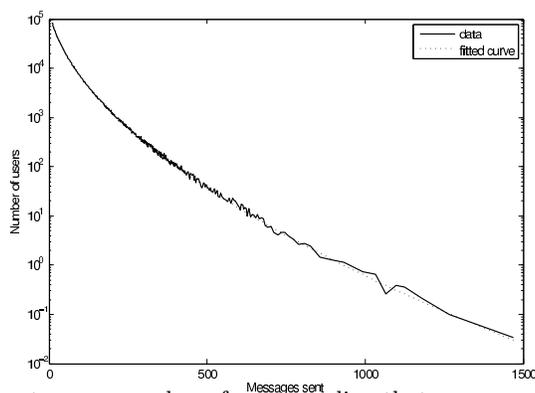} }
	\caption{Number of messages sent versus number of users sending that many messages (solid line) and fitted curve (dashed)}
	\label{f:powerpower}
\end{figure}

The number of messages sent per user in our data is shown in figure \ref{f:powerpower}. In the figure, the data has been binned so that each bin contains a non-trivial number of counts.
The data is remarkably well described by the curve $An^{-\alpha n^{\beta}}$ with $\alpha = 0.0947$, $\beta = 0.426$, and properly normalized $A$. In fact, a log-likelihood test against the observed data (binned so that no bin's count is less than 10) yields a $p$-value of 0.52 when applied between $n=10$ and the upper limit, and a $p$-value of 0.047 when applied between $n=100$ and the upper limit. 

It should be noted that the probability distribution
\[
 P(x) \sim x^{-\alpha x^{\beta}},~~~ \alpha>0, ~~~0<\beta<1 ,
\] 
which has received little attention, provides a middle ground between the heavy-tail Pareto or power-law and thin-tailed exponential distributions in terms of its asymptotic behavior. Its moments are all well-defined\footnote{Indeed, for any moment $m$ and constant $c>1$, the function satisfies
\[
P(x)x^m = x^{m - \alpha x^\beta} < x^{-c}
\] 
for all $x > (\frac{m+c}{\alpha})^{1/\beta} $, and the integral converges.} although its mode lies at $e^{-1/\beta} < 1$ and is thus inapplicable in the case of integer counts. Of course, the Pareto distribution is recovered when $\beta=0$. The mechanism underlying this distribution's excellent description of our data is unclear and provides a subject for future work.

\section{Mathematical details} \label{s:C}

\subsection{Similarity of schools' weekly distributions}\label{s:C1}
This section justifies the statement that the messages sent by users at individual schools, or the messages sent over a limited range of dates, displayed ``the same'' weekly pattern as that of the aggregate data (figure \ref{f:week}).  Two approaches may be taken to do this. 

First, for a given school, we considered the percentage of messages sent for each hour of the week to be a random variable. Since there are 168 hours in a week, we thus have 168 random variables, each with expected value 1/168. Of course, each hour displays some variation from this average. We then obtained a second series of random variables from the counts for second school, or from the overall aggregate data. Then, we computed the correlation coefficient as in equations \ref{e:corr1} and \ref{e:corr2} below, where however the sum runs over all hours of the week, the variables $x_i$ represent the percentage for school $x$ in hour $i$, and the averages are all 1/168. The resulting correlation coefficients are very strongly positive for all pairs of schools and for various time slices.  In this statistical sense, the patterns are similar. 

A second approach is to attempt to calculate a ``reference hour'' which aligns a particular school's pattern to the overall pattern\footnote{This is precisely the challenge of determining a time zone for each school, although in practice this was more difficult since the overall pattern was hard to determine before the time zones were known. Our approach was to calculate an overall average, make a best guess for each school, and repeat until a stable solution was found.}. It was observed that the log likelihood function overwhelmingly favored one particular hour as the reference hour, and that this was precisely the hour which produced the largest correlation coefficients described in the previous paragraph. The ease with which the reference hour is determined in this fashion is a second justification of the statement that individual schools had the same weekly pattern taken separately as the aggregate data. 

\subsection{Correlation coefficient}\label{s:C2}
To compare variations in data and search for correlations, we used the following standard methodology \cite{Rice}. The explanation will be given in terms of the correlations described in section XX and by figures \ref{f:r3} and \ref{f:r}, although a similar (but subtly different) approach was used in the test of appendix \ref{s:C1}.

The data in question is percentages of messages plus pokes sent per hour of the week, grouped by school. For a given hour, we consider the percentage of counts for each school to be a random variable. For each hour of the week, we thus have a sample of 496 draws, one for each school. The estimated mean for the distribution for hour $x$ is obtained by averaging the percentages for that hour over all 496 schools, while the variance is estimated by
$ s_x = \frac{1}{n}\sum(x_i - \langle x \rangle)^2,$ where $x_i$ is the percentage of messages for school $i$ during hour $x$ of the week, $\langle x \rangle$ is the estimated mean percentage for that hour, and the sum runs over all 496 schools. Next, we estimate the covariance by hour,
\begin{equation}\label{e:corr1}
s_{xy} = \frac{1}{n}\sum_{i=1}^{496}(x_i - \langle x \rangle)(y_i - \langle y \rangle),
\end{equation}
and thereby obtain the estimated covariance correlation between hours $x$ and $y$:
\begin{equation}\label{e:corr2}
r_{xy} = \frac{s_{xy}}{\sqrt{s_xs_y}}.
\end{equation}
$r_{xy}$ lies between --1 and 1 and measures the degree to which schools' deviations from the average for hours $x$ and $y$ are correlated.

\bibliographystyle{plain}
\footnotesize
\bibliography{fb_arxiv}
\end{document}